\documentclass[a4paper,11pt]{article}
\usepackage{amsmath}
\usepackage{amsfonts}
\usepackage{amssymb}
\usepackage{bm}
\usepackage[utf8x]{inputenc}
\usepackage{ae}
\usepackage[T1]{fontenc}
\usepackage[english]{babel}
\usepackage{graphicx}
\newcommand{\indep}{\rotatebox[origin=c]{90}{$\models$}}
\usepackage{makecell}
\usepackage{placeins}
\usepackage{tabularx}
\usepackage{tabulary}
\usepackage{setspace}
\usepackage{tablefootnote}
\usepackage[flushmargin]{footmisc}
\usepackage[comma]{natbib}
\usepackage{array}
\usepackage{float}
\usepackage{caption}
\usepackage{tikz}
\usetikzlibrary{shapes.geometric}
\usepackage[a4paper,left=2.5cm,right=2.5cm,top=3cm,bottom=3cm]{geometry}
\usepackage[colorinlistoftodos]{todonotes}
\usepackage{pdfpages}
\usepackage{longtable}
\usepackage{adjustbox}
\usepackage{booktabs}
\usepackage{multirow}
\usepackage{lscape}
\usepackage[colorlinks=true, allcolors=blue]{hyperref}

\usepackage[flushleft]{threeparttable}
\usepackage{inputenc}



\begin{document} \doublespacing \pagestyle{plain}
	
	\def\ci{\perp\!\!\!\perp}
	\begin{center}
		
		{\LARGE Free public transport to the destination: A causal analysis of tourists' travel mode choice}

		{\large \vspace{0.8cm}}
		
		{\large Kevin Blättler, Hannes Wallimann and Widar von Arx }\medskip

		{\small {University of Applied Sciences and Arts Lucerne, Institute of Tourism and Mobility} \bigskip }
		
		{\large \vspace{0.8cm}}
		
		{\large Version February 2024}\medskip

	\end{center}
	
	\smallskip

	\noindent \textbf{Abstract:} {
		In this paper, we assess the impact of a fare-free public transport policy for overnight guests on travel mode choice to a Swiss tourism destination. The policy directly targets domestic transport to and from a destination, the substantial contributor to the CO$_2$ emissions of overnight trips. Based on a survey sample, we identify the effect with the help of the random element that the information on the offer from a hotelier to the guest varies in day-to-day business. We estimate a shift from private cars to public transport due to the policy of, on average, 14.8 and 11.6 percentage points, depending on the application of propensity score matching and causal forest. This knowledge is relevant for policy-makers to design future offers that include more sustainable travels to a destination. Overall, our paper exemplifies how such an effect of comparable natural experiments in the travel and tourism industry can be properly identified with a causal framework and underlying assumptions. 
	}
	
	{\small \smallskip }
	{\small \smallskip }
	{\small \smallskip }
	
	
	{\small \smallskip }
	{\small \smallskip }
	{\small \smallskip }
	
	{\small \noindent \textbf{Acknowledgments:} We are grateful to the SBB Research Fund for financial support. We have benefited from comments by Guido Buob, Martin Huber, Michael Stiebe, Vu Thi Thao, Noah Balthasar, Christian Laesser, and Kimberly Lippuner.}
	
	\bigskip
	\bigskip
	\bigskip
	\bigskip
	
	{\small {\scriptsize 
\begin{spacing}{1.5}\noindent  
\textbf{Addresses for correspondence:} Kevin Blättler, University of Applied Sciences and Arts Lucerne, Rösslimatte 48, 6002 Lucerne, \href{mailto:kevin.blaettler@hslu.ch}{kevin.blaettler@hslu.ch}; Hannes Wallimann, \href{mailto:hannes.wallimann@hslu.ch}{hannes.wallimann@hslu.ch}; Widar von Arx, \href{mailto:widar.vonarx@hslu.ch}{widar.vonarx@hslu.ch}.
\end{spacing}
			
		}\thispagestyle{empty}\pagebreak  }

	{\small \renewcommand{\thefootnote}{\arabic{footnote}} %
		\setcounter{footnote}{0}  \pagebreak \setcounter{footnote}{0} \pagebreak %
		\setcounter{page}{1} }
	
\section{Introduction}\label{introduction}

Tourism's global carbon footprint accounts for about 8\% of global greenhouse gas emissions \citep{lenzen2018carbon}. Transportation contributes 72\% substantially to the global CO$_2$ emissions of overnight tourism \citep{peeters2010tourism}. Whereas in international travels, most CO$_2$ emissions stem from air travel, emissions from private car trips gain importance in domestic overnight stays. Even though private car usage emits more CO$_2$ per passenger kilometre than public transport, \citet{peeters2010tourism} estimate that 90\% of domestic trips in developed countries are made by car. Since a shift towards public transport helps decrease CO$_2$ emissions, there exists a wide range of studies discussing the mode shift from private cars towards public transport \citep[][]{redman2013quality}. In the context of tourism, policies that effectively incentivize leisure travelers to use public transport instead of private cars (or airplanes) are at the forefront of the thinking of researchers and policy-makers \citep[][]{le2015tourist}.

However, besides its tremendous impact on the environment, there is limited information on such natural experiment estimates, where policies directly target the arrival and departure of overnight tourists---which differ from other travelers, e.g., by traveling with more luggage. With the prospect of considerably reducing CO$_2$ emissions, a Swiss tourism destination launched an innovative offer, where overnight guests who stay for at least three nights can order a free public transport ticket (for the whole Swiss public transport network) for their arrival and departure---reducing the monetary cost for sustainable arrivals and departures to zero. Adding to studies investigating fare-free policies \citep[see, e.g., ][]{cats2017prospects,vstraub2023belchatow,lu2024analyzing}, this paper analyzes the effect of a free arrival and departure offer on the travel mode choice of overnight tourists. Whereas research papers present various estimates of the effects of price changes in public transport due to natural experiments \citep[see, e.g., ][]{kholodov2021public,wallimann2023price}, studies identifying the effects of natural policy experiments on overnight travelers are rare. Moreover, our study investigating domestic travel is valuable insofar as the tourism literature on travel mode choices discusses mainly international travels \citep[see, e.g., ][]{thrane2015examining}.

In our case, the free arrival and departure offer to and from the destination is only valid when guests actively order the public transport ticket before arrival. Our causal analysis takes advantage of the random element that the information on the offer from the hotelier to the guest varies in day-to-day business. Therefore, we can split the guests with regard to the information status into two groups, i.e., informed and non-informed guests. Using matching methods (i.e., causal forest \citep{athey2019generalized} and propensity score matching \citep[][]{rosenbaum1983central}) and based on the so-called "selection-on-observable assumption", we answer the research question on the causal effect of the free arrival and departure offer on mode shift from private car to public transport among overnight guests. Finally, the thing to notice is that to identify our theoretical mechanism of interest---i.e., the causal effect of the free arrival and departure offer on mode shift from private car to public transportation---we estimate the effect (only) among guests, not being aware of the offer during the booking process.

We obtain average treatment effects (ATE) of 0.116 and 0.148, both being statistically significant at conventional levels, when applying the causal forest and propensity score matching, respectively. That means when a guest gets informed by the hotelier, the probability that the guest travels by public transportation (instead of a car) increases by 11.6 or 14.8 percentage points (depending on the statistical method). We benchmark our estimates in several robustness checks, e.g., by using guests in surrounding regions without such an offer as a control group. These investigations show that the effect remains significantly positive. To sum up, we provide the first empirical evidence that a free arrival and departure offer for overnight tourists can effectively shift trips to the destination from private car to public transportation.

The remainder of the paper is organized as follows. Section \ref{Litrev} discusses the relevant literature. Section \ref{background} contains the background of the offer in Switzerland and data steaming from a unique survey in the region of interest. In Section \ref{Identification_estimation}, we describe how we identify the causal effects. Section \ref{Desc} outlines descriptive statistics. In Section \ref{Results}, we show the estimated effects of the free arrival and departure offer on mode shift. Section \ref{Discussion} discusses the results in the practical and political context. Finally, Section \ref{Conclusion} concludes.

\section{Literature review}\label{Litrev}

Our study relates to the literature on fare-free policies, the most drastic possible price reduction, as we generate new insights for researchers and policy-makers by analyzing a free public transport policy for the customer segment of overnight guests. An example is the paper of \cite{cats2017prospects}, concluding that fare-free public transport in Tallinn (Estonia) led to a demand increase (i.e., number of trips) of 14\%, while in the rest of the country during the period of investigation, the mode share of public transport decreased. On the other hand, analyzed offers of free public transport exist for a specific customer segment. Based on a case of students from Brussels (Belgium), \cite{de2006impact} show an increase in public transport usage among students benefiting from the offer. \cite{rotaris2014impact} conclude, based on a case of the University of Trieste (Italy), that fully subsidizing buses would raise bus share from 53\% to 61-81\%. \cite{shin2021exploring} estimates there was a 16\% increase in subway use by citizens aged 65 and above after a fare-free policy was introduced for this age group in Seoul. Recently, \cite{vstraub2023belchatow} investigate 93 municipalities engaged in fare-free programs and show, for instance, that these programs are more likely to emerge in localities with stable and increasing populations and relatively high levels of public expenditure. However, in contrast to these studies, we focus on the travel mode choices of tourists with overnight stays. 

Moreover, the thing to notice is that studies using quasi-experimental approaches (such as the so-called selection-on-observables assumption as in our study) to investigate the effect on policies on guests' travel mode choices are rare, where there exist estimates of the effects of price changes in public transport due to natural experiments \citep[see, e.g., ][]{hoang2021impacts,kholodov2021public,wallimann2023price}. Recently, closely related to our study, \citet{andersson2023complexity} investigate the effect of a free public transport (PT) card intervention on mode shift using a quasi-experimental setting. As in our paper, the latter study examines the influence of measures on a travel mode shift and not rarely an increase in the number of travelers on public transport \citep[as, e.g.,][]{wallimann2023price}. Besides fare policies, quasi-experimental approaches are also applied to estimate mode shift effects of public transport quality improvements such as improved accessibility \citep[as, e.g.,][]{dai2020effects,wang2023effect}. However, again, we differ in that we do this for a specific segment---the overnight guests.

The mean of transport of overnight tourism is mainly analyzed for long-distance---international---travels. \cite{thrane2015examining} shows that distance matters for the travel mode choice, as the probability of choosing air transportation over private and public transportation increases significantly with longer routes. The results suggest a turning point at around 400 km at which tourists shift from using private cars or public transportation to using air transportation. Compared to this literature, our paper focuses mainly on domestic trips, for which private cars and public transportation are the major counterparts. \cite{pellegrini2021relationship} examine the travel mode choice to reach the destination for domestic trips in Switzerland and highlight that the trip-related decisions such as length of stay, mean of transport, and accommodation type correlate. They also show that solo travelers are more likely to travel with public transport than groups, indicating that the costs of traveling by different means of transport vary for different group sizes. \cite{masiero2013tourists}, also investigating Swiss tourism, find that travel mode choice and movement patterns during holidays are interlinked. Additionally, another stream of literature related to the underlying study focuses on the mobility behavior at the destination rather than the travel mode choice to reach a destination. For instance, \cite{bursa2022intra} and \cite{bursa2022travel} suggest that, inter alia, travel time, group composition, trip purpose, weather, and information about the destination are associated with the mode choice for activities within a destination. \cite{zamparini2021sustainable} add that besides the mobility at home and mobility patterns within the destination, the transport mode to reach a destination relates to the mobility behavior within a destination. 

Moreover, when discussing determinants of tourists' travel mode choices, the influence of public transport supply is crucial. For instance, \cite{gronau2007key} argue that the destination's target groups should have a proneness towards public transport, such that public transport policies can be effective. However, if this prerequisite is given, quality improvement has the potential to shift towards public transport. Therefore, \cite{le2015tourist} state that tourists rather use public transport in urban areas more than in remote areas since urban transport systems are typically of higher quality. \cite{pagliara2017exploring} find that improvements in connectivity and accessibility in public transportation in Italy increase demand for the destination. The complementary study of \cite{boto2023effect} observes that public transport improvements in Spain increased the share of arrivals in the low season, indicating a modal shift. However, \cite{bursa2022travel} argue that price interventions neither for public transportation nor private cars induce a substantial shift to public transportation. Therefore, \cite{orsi2014assessing} summarize that effective policies should cautiously combine public transport policies and car-use regulations. Finally, \cite{romao2021determinants} point out that public transport services can increase the overall trip satisfaction of tourists.

In a broader picture, our paper has implications for tourism destination management, as the offer might have positive spillover effects on accommodation businesses. For instance, \cite{wallimann2022complementary} shows that drastic price reductions of skiing passes positively affected the number of overnight stays in a Swiss destination. With the offer at hand exclusively valid for guests that stay at least three nights, the offer also targets particular guests that generate (per arrival) above-average economic impact on a destination \citep{de2008determinants}. On the other hand, the average environmental impact of tourists (per day) decreases, which might lead to a more sustainable tourist mix in the destination. For a discussion about the optimal length of stay regarding earnings and CO$_2$ emissions, see, e.g., \cite{oklevik2020overtourism}.

\section{Background and survey}\label{background}

Our study focuses on Switzerland, a country in the middle of Europe where tourism generates 16.8 billion Swiss francs gross value added \citep{STV2023} and contributes about 3\% to Swiss GDP \citep{regiosuisse2023Monitoring}. Approximately 4\% of the Swiss export revenue stems from tourism, and about 3.8\% of all employees in Switzerland work in the tourism industry \citep{STV2023}. The Swiss resident population undertook 16.3 million trips with one or more overnight stays, of which 9.1 million were within Switzerland \citep{STV2023}. The public transport system in Switzerland, due to the high level of system integration with frequent services, comprehensive fare integration, and synchronized timetables, is of high quality of service \citep[see, e.g.,][]{thao2020swiss}. However, of those Swiss residents with overnight stays traveling within Switzerland, 57.1\% travel by car to their destination in Switzerland, compared to (only) 31.9\% traveling by train according to \citet{TourismMonitor2017}. 

Our area of interest is the Swiss canton Appenzell Innerrhoden, a small, rural canton located in the East Alpine region of Switzerland. With 16,000 inhabitants, it is the least populous canton in Switzerland.\footnote{Officially, it is a so-called half-canton, not being relevant for our study.} The canton is well known for its main town, Appenzell, and the surrounding nature and mountains, as well as its cultural heritage. Hence, tourism contributes 12.8\% to the cantonal GDP, and a considerable share of 16.8\% of inhabitants work in tourist-related businesses \citep{Wertschoepfung2019Appenzell}. The main town is accessible by train at a half-hour frequency from the Swiss cities of Herisau and St.Gallen. Most smaller towns are also accessible by these train lines, with no or one changeover in Appenzell.

The free arrival and departure offer was launched by the destination marketing organization (henceforth also referred to as DMO) in 2020. Since then, overnight guests who stay for at least three nights in the canton of Appenzell Innerrhoden can order a free public transport ticket (for the whole Swiss public transport network) for their arrival and departure. During the pilot phase from 2020 to 2022, the offer was co-financed by the New Regional Policy, which aims to reduce regional disparities by financially supporting innovative projects and initiatives in rural regions \citep{Geschaeftsbericht2020AI}. The seed capital provided in the framework of the New Regional Policy is paid by the federal government and the respective canton in equal parts.\footnote{See \href{https://regiosuisse.ch/en/new-regional-policy-nrp}{https://regiosuisse.ch/en/new-regional-policy-nrp} (accessed on October 18, 2023).} Since 2023, the DMO has independently financed and promoted the offer. The accommodation businesses do not have to make a direct contribution.

Appenzell Innerrhoden has also established a "guest card" for a couple of decades that permits guests who stay at least three nights in an accommodation to use 20 attractions and public transport within the destination free of charge.\footnote{See \href{https://www.appenzell.ch/de/unterkunft/appenzeller-ferienkarte.html}{https://www.appenzell.ch/de/unterkunft/appenzeller-ferienkarte.html} (accessed on October 18, 2023). The card is valid for a maximum of seven nights. Guests who stay longer than seven nights receive a new Appenzell guest card free of charge.} While public transport free of charge within the destination during the stay is implemented in various Swiss (and German-speaking) tourism destinations (see, e.g., \cite{gronau2017encouraging}), public transport free of charge for the arrival and departure to and from the destination on top of it---our policy of interest---is novel.\footnote{As far as we know, besides the offer in Appenzell Innerrhoden, there exist only a few smaller-scale free arrival and departure offers in Switzerland. They are either limited to certain hotels (e.g., Glarnerland) or specific activities (e.g., Nature Parks).}  However, in contrast to the free arrival and departure offer, the accommodation businesses co-finance the guest card \citep{Geschaeftsbericht2022AI}. In our study period, the free arrival and departure offer was used by 2,373 overnight guests, and 12,886 guest cards were distributed. 

To gather the relevant data, we conducted an online survey between May and October 2023 based on the software Unipark. Our leading partner in carrying out the survey was the Appenzell Innerrhoden DMO. In cooperation with the DMO, we addressed 4,333 guests owning a guest card by mail. Guests were directed to the online survey via a link. Additionally, we attached the link to the free arrival and departure offer, and the hotels distributed flyers with a QR-code to the online survey among their overnight guests.\footnote{The flyers made it possible also to reach guests staying in the surrounding tourism destination Appenzell Ausserrhoden and Toggenburg.} 1,871 guests that stayed at least three nights in Appenzell Innerrhoden completed the survey.

Retrieved from the literature (see Section \ref{Litrev}), we asked questions about the travel mode choice and all factors influencing this decision. After collecting the data about the mode choice and the explanatory variables, we also questioned the overnight guests about the free arrival and departure offer. For our analysis, it is fundamental to determine whether and when they received the information about the free arrival and departure offer in this part. That is because this knowledge allows us to identify the effect of the policy on travel mode choice, which we explain in the forthcoming section in greater detail.

\section{Identification and estimation}\label{Identification_estimation}

\subsection{Definition of causal effects}

\begin{figure}[H] 
	\includegraphics[scale=.5]{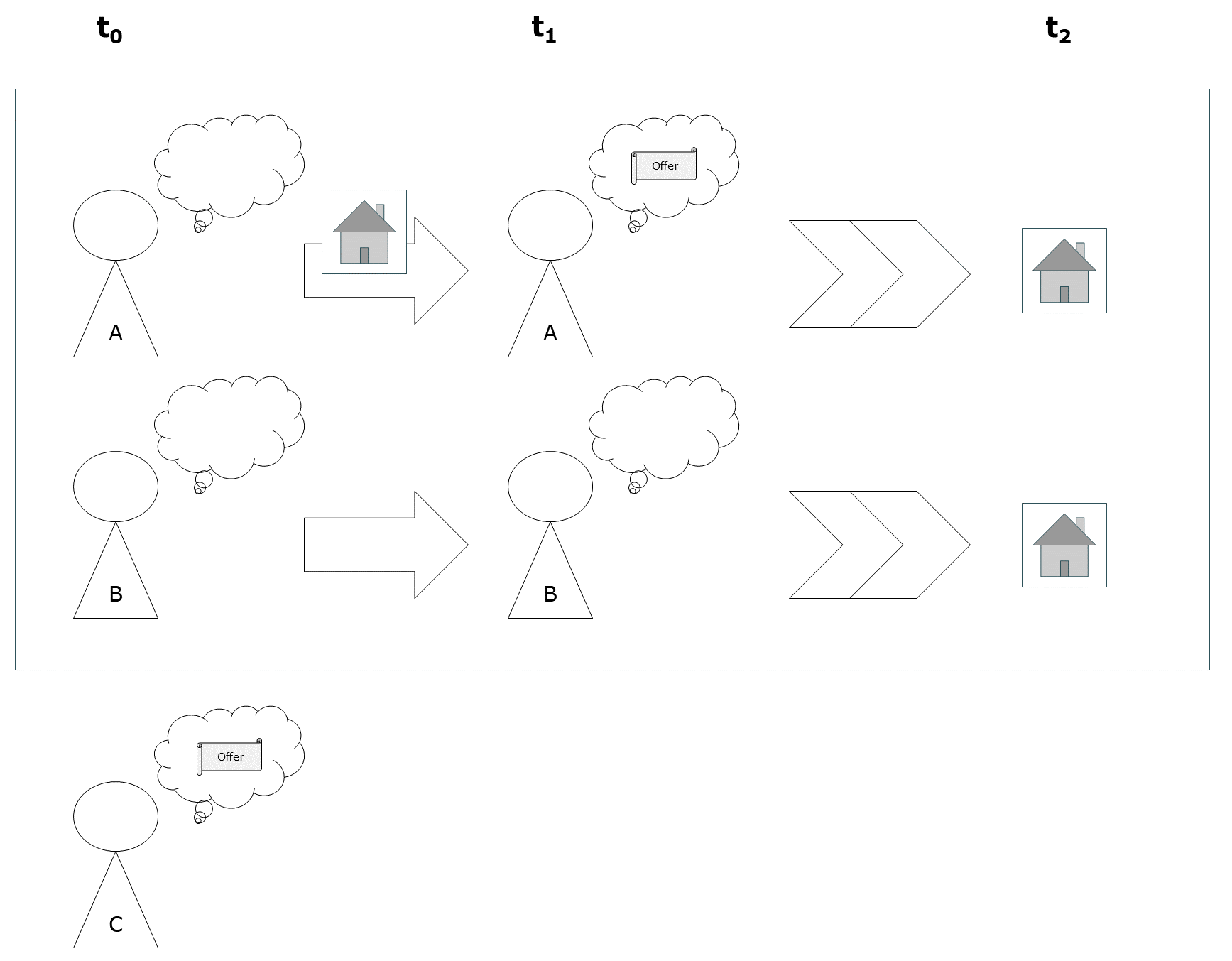}
	\centering \caption{Information status of guests at three time stages} \label{Fig:Defcausal}
\end{figure}

Our causal analysis is based on the potential outcome framework \citep[see, for instance,][]{rubin1974estimating}: The causal effect of a treatment is the difference between the outcomes of individuals to a certain point in time exposed and not exposed to a treatment initiated at an earlier stage. At time $t_1$, the hotelier may (guest "A" in Figure \ref{Fig:Defcausal}) or may not (guest "B" in Figure \ref{Fig:Defcausal}) inform these guests. Therefore, we take advantage of the fact that the information from the hotelier to the guest varies due to everyday stress and duties. Hence, we define $D$ as a binary treatment indicator, whether the accommodation informs the guest about the offer ($D=1$, i.e., guest "A" in Figure \ref{Fig:Defcausal}) or the accommodation does not inform the guest ($D=0$, i.e., guest "B" in Figure \ref{Fig:Defcausal}). In $t_2$, we either observe potential outcome $Y(1)$ or $Y(0)$, whereas the observable outcome $Y$ is “travel mode choice”---$Y=1$ and $Y=0$ indicate public transport and no public transport usage, respectively (depicted with the two chevron arrows to the accommodation in Figure \ref{Fig:Defcausal}).\footnote{Note that negligible 1\% of the guests travel with the bike to the destination.} Using the rhetoric of causal inference, we can uncover the average causal effect---also known as average treatment effect (ATE)---of the information that one could arrive and departure with public transport free of charge on the outcome travel mode choice at time $t_2$. The ATE corresponds to the difference in the average potential outcomes $Y(1)$ and $Y(0)$ in the population of interest:

\begin{equation}
	\Delta=E[Y(1)]-E[Y(0)].
\end{equation}

To ensure the identification of the effect, we only look at those guests who were not aware of the free arrival and departure offer at time $t_0$ of the booking process. On the other hand, guest "C" is already informed about the offer when booking the holidays (e.g., because of marketing). As these such guests (represented by "C") who use public transportation may differ in terms of unobservable characteristics from guests who were not aware of the free arrival and departure offer at the time of the booking process (represented by "A" and "B"), we, for our causal analysis, ignore them.

As with many empirical applications, our analysis relies on observational (nonrandomized) data. Therefore, we uncover the treatment effect with the "selection-on-observable assumption". The idea is to compare the outcomes of individuals exposed and not exposed to the treatment that are similar in terms of covariates---characteristics that jointly influence both the decision to receive treatment and the outcome of interest \citep[see, e.g., ][]{huber2023causal}. Therefore, we assume that by controlling for observed characteristics, the treatment is as good as if it were randomly assigned among those treated and non-treated subjects (as in an experiment). Put differently, we can avoid that the treatment effect is mixed up with any impact of differences in covariates and interpret differences in the outcomes to be exclusively caused by differences in the treatment.

\begin{figure}[H]
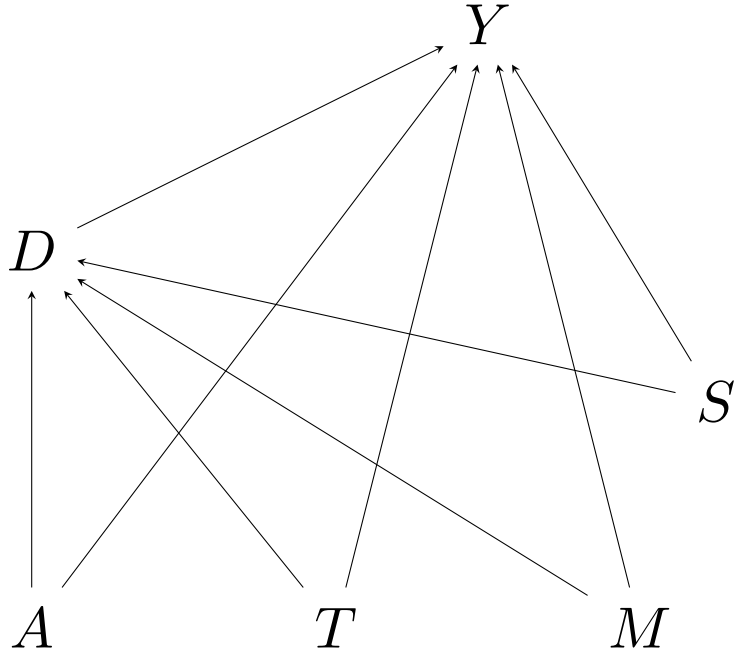

	\centering \caption{\label{figuresetup}  Causal framework}\bigskip
	\definecolor{offwhite}{HTML}{F2EDED} 
	\tikzset{> = stealth} 
	\tikz{ 
		\node[black, scale=2] (d) at (0,5) {$D$}; 
		\node[black, scale=2] (t) at (0,0) {$A$}; 
		\node[black, scale=2] (s) at (4,0) {$T$}; 
		\node[black, scale=2] (m) at (8,0) {$M$};
		\node[black, scale=2] (a) at (9,3) {$S$};
		\node[black, scale=2] (y) at (6,8) {$Y$}; 
		\path[->, draw=black] (d) edge (y); 
		\path[->, draw=black] (t) edge (y); 
		\path[->, draw=black] (s) edge (y); 
		\path[->, draw=black] (m) edge (y);
		\path[->, draw=black] (t) edge (d); 
		\path[->, draw=black] (s) edge (d); 
		\path[->, draw=black] (m) edge (d);
		\path[->, draw=black] (a) edge (d);
		\path[->, draw=black] (a) edge (y);
		
	}
\end{figure}

The directed acyclic graph (DAG) in Figure \ref{figuresetup} illustrates the causal framework of our identification strategy. Our entire set of observed characteristics $X$ can be subsumed under accommodation-specific characteristics $A$, trip-related characteristics $T$, mobility tools $M$, and socio-demographic characteristics $S$. For the accommodation-specific characteristics $(A)$, we include a hotel-specific ratio of informed vs. uninformed guests per accommodation to account for the probability that a guest is informed by the different hoteliers as well as two dummy variables for the accommodation type and the accessibility by train \citep[see for the latter two variables, e.g., ][]{pagliara2017exploring,pellegrini2021relationship}. Further, we add the relevant trip-related characteristics $(T)$ travel party composition, travel purpose, length of stay, distance with the private car from home to the accommodation, travel time difference between car and public transport, and Swiss residence \citep[see, e.g., ][]{rodriguez2018length}. Moreover, as mobility tool ownership $(M)$ influences the travel behavior \citep[see, e.g., ][]{thao2020impact}, we consider the covariates accounting for car and public transport season ticket ownership. Finally, we also control for socio-demographic characteristics $S$ as age, income, and gender \citep[see, e.g., ][]{rodriguez2018length,thao2020impact}. 

To ensure the identification of the effect, we drop two particular subgroups that cannot, or only to a limited extent, gain benefit from the free arrival and departure offer. On the one hand, we ignore guests with a GA Travelcard. This season ticket allows the unlimited use of public transport Swiss-wide and therefore yields the same benefit as the offer of interest in this paper. On the other hand, we omit guests with an arrival journey to the destination that is longer than 400 km. From this threshold, air transportation becomes relevant (\cite{thrane2015examining}).\footnote{The most distant city in Switzerland, Geneva, is less than 400 km away (by car and public transport) from the canton of Appenzell Innerrhoden.} The thing to notice is that our subset of observations now differs from the typical average treatment effect estimand ATE over the full population (using the rhetoric of causal inference, this procedure is referred to as "moving the goalpost", see, e.g., \citet{crump2006moving}).

\subsection{Identifying assumptions} \label{Assumptions}

Identifying the potential outcomes under treatment and non-treatment relies on assumptions how the real world works. Therefore, our estimations of the effect of the free travel offer also rely on assumptions. 

\textbf{Assumption 1 (conditional independence assumption):} \newline
Assumption 1 (also called selection on observable) is satisfied when the potential outcomes ($Y(1)$,$Y(0)$) are conditionally independent of the treatment ($D$) when controlling for covariates ($X$), formally

\begin{equation}
	\{Y(1),Y(0)\} \indep D|X. 
\end{equation}

This assumption holds, when all covariates that jointly influence potential outcomes and treatment are observed and controlled for. This implies that the treatment is as good as randomly assigned among treated and non-treated overnight guests with the same characteristics. Due to our rich set of observed characteristics $X$---subsumed under accommodation-specific characteristics $A$, trip-related characteristics $T$, mobility tools $M$, and socio-demographic characteristics $S$---that we derived from the transport literature, it is realstic that the conditional independence assumption is fulfilled.

\textbf{Assumption 2 (common support):} \newline
Assumption 2 states that the conditional treatment probability is larger than zero and smaller than one such that ($D$) is not deterministic in $X$, formally:

\begin{equation}
	0<p(X)<1, 
\end{equation}

where $p(X)=Pr(D=1|X)$ is the conditional treatment probability---also called the propensity score. In other words, when comparing the treatment and control groups, there must be substantial overlap in the distribution of the observed covariates. An example in our case is that for every value of car ownership (i.e., "Yes" and "No"), there must be subjects receiving and not receiving the information about free arrival and departure offer.

\textbf{Assumption 3 (exogeneity):} \newline
Assumption 3 stipulates that $X$ is not a function of $D$ and therefore does not contain characteristics that are affected by the treatment, formally:

\begin{equation}
	X(1) =X(0)=X.
\end{equation}

Therefore, we leave out overnight guests adjusting their number of nights after learning about the offer from one and two nights to three nights, as for those, the covariate "length of stay" is effected by the treatment. 

\textbf{Assumption 4 (random sample assumption):} \newline
Assumption 4 is satisfied when our collected data is a random sample condition on $X$ from the population of interest. To check whether we identify our theoretical mechanism of interest (and not an estimate due to the vast number of guest card owners in our sample being different from the overall population), we compare the CATEs derived from the model derived from the sample of (only) guest card owners and those from the overall population.

\textbf{Assumption 5 (identifiability of treatment status at $t_1$):} \newline
By Assumption 5, we assume we know whether a person was informed by the hotel ($D=1$) or not ($D=0$). Assumption 5 is satisfied in the absence of misreporting regarding this information. 

Following \cite{huber2023causal} to show how our assumptions permit identifying the average treatment effect (ATE), let us use $\mu_d(x)=E[Y|D=d,X=x]$ to denote the expected conditional mean outcome given the binary treatment $D$ (i.e., "information status") which is either $0$ ("not informed") or $1$ ("informed"), and the observed covariates $X$ (including accommodation-specific characteristics $A$, trip-related characteristics $T$, mobility tools $M$, and socio-demographic characteristics $S$). $\mu_1(x)-\mu_0(x)$ identifies the causal effect among individuals which share the same values $x$ of the observed covariates $X$. We denote the average effect under the condition that subjects share the same covariate values X = x as conditional average treatment effect (CATE):

\begin{equation}
	\Delta_x=E[Y(1)|X=x]-E[Y(0)|X=x]=\mu_1(x)-\mu_0(x).
\end{equation}

Averaging across all CATEs among all values of $x$ which the covariates $X$ take in the population permits to identify the average treatment effect (ATE): 

\begin{equation}
	\Delta=E[\mu_1(X)-\mu_0(X)].
\end{equation}

\subsection{Estimation based on propensity score matching and causal forest} 

In our study, we use matching methods to derive our average treatment effect estimands. To make treatment and control groups comparable, matching methods create a set of weights for each observation. To estimate the treatment’s effect, we can calculate a weighted mean of the outcomes. As matching variables, we use accommodation-specific characteristics $A$, trip-related characteristics $T$, mobility tools $M$, and socio-demographic characteristics $S$ (see also Figure \ref{figuresetup}). Statistically speaking, matching is the process of closing back doors between the treatment variable $D$ and the outcome variable $Y$ \citep[see, e.g., ][]{huntington2021effect}.

\citet{rosenbaum1983central} demonstrate that conditioning on the propensity score $p(X)$ balances the distribution of X across the treatment group and control group such that the covariates are potential outcomes are conditionally independent of the treatment conditional on the propensity score: X\indep D|p(X). In Figure \ref{figuresetup_PSM} in Appendix \ref{Appendix_Causal_Framework_PSM}, we see that the propensity score can be interpreted as a function of our covariates $X$ through which any effect of X on D operates. Therefore, we can identify the ATE when controlling for the propensity score $p(X)$ as: 

\begin{equation}
	\Delta_x=E[\mu_1(p(x))-\mu_0(p(x))]. 
\end{equation}

To achieve the effect through propensity score matching, we use logit regression to estimate the propensity scores. To account for the estimation based on propensity score, we calculate and display bootstrap-based standard errors.

We also apply the causal forest approach of \citet{wager2018estimation} and \citet{athey2019generalized}; see also \cite{huber2022business} for the first application of causal machine learning in the public transportation literature. The causal forest approach estimates propensity scores and ATEs using random forest. Causal forest is especially useful in the presence of irrelevant covariates. Also, the causal forest has the nice properties to estimate effect heterogeneity, the CATEs. Both strengths enable us to analyze the case more flexibly.

To estimate the causal effects of the free arrival and departure offer with propensity score matching and causal forest, we use \textit{Matching} and \textit{grf} packages in the statistical software \textsf{R}.

\section{Descriptive analysis}\label{Desc}

Based on our identification,  it remains a sample with 843 observations. 189 observations have missing values, from which 157 have only one covariate missing.\footnote{Missing variables exist for the covariates \textit{Age}, \textit{Gender}, \textit{High income}, \textit{Car ownership} and \textit{Travel time difference}.} Descriptive statistics suggests that these covariates are missing at random (see Table \ref{table:desc_imputation} in Appendix \ref{Appendix_A}), and hence, we decide to drop those variables. However, we impute the missing values in a robustness check and re-estimate the effect (see Section \ref{Result:Robust}).

In Table \ref{table:desc}, we present descriptive statistics of our set of matching variables ($X$) and the outcome variable ($Y$) by the binary indicator $D$ taking the value $D=1$ for informed guests. Different hoteliers prioritize the guest information in their day-to-day business and accordingly make guests more or less aware of the offer during the booking process. Therefore, we observe in Table \ref{table:desc} that the hotel-specific ratio of informed vs. uninformed guests varies between treatment ($D=1$) and control group ($D=0$). 

Naturally, the hotel-specific ratio of informed guests is higher in the treatment group than in the control group, as this variable reflects the varying probability of hoteliers informing their guests (note that for this variable, we consider all holiday flats as a hotel). Also, uninformed guests stay on average more in holiday flats than in hotels. Hotel guest information upon arrival might be more professional and standardized than that of holiday flats. Moreover, guests in the treatment group rather stay in accommodations accessible by train, whereas more guests in the control groups stay in accommodations only accessible by bus. This is possible because hotels that are directly accessible by train might expect a higher benefit from promoting the offer. As expected, the accommodation-specific variables $A$ vary, reflecting the varying frequency of each hotelier informing their overnight guests during the booking process. 

Most guests do not travel with their families, whereas the guest's primary holiday purpose is nature or hiking in both groups, with both proportions being slightly higher in the control group. Other trip-related variables ($T$), such as the length of stay, swiss residence, travel distance in km by car, and travel time difference between car and public transport usage are comparable for both groups. The two latter we accessed on Google Maps knowing the anonymized destination and origin (postal code) of guests.\footnote{See \href{https://console.cloud.google.com/google/maps-apis}{https://console.cloud.google.com/google/maps-apis}, accessed on November 11, 2023.}

Furthermore, more guests in the treatment group than in the control group own a Half Fare Travelcard (implying a price reduction of 50\% for public transport tickets in Switzerland). This difference can be interpreted as guests with a Half Fare Travelcard being more prone to use public transport and, therefore, instead asking the hotelier for public transport offers during the booking process. However, as a thing to notice, this does not directly imply that more people with a Half Fare Travelcard actually use the offer as our treatment is the information about the offer and not the offer itself. Secondly, the treatment and control groups are similar in car ownership, whereas the share of people owning a car is high in our sample. Finally, age, gender, and income status (i.e., represented by a guest's household owning more than 12,000 Swiss francs) are similarly spread among treatment and control groups.

Our outcome variable states whether a guest uses public transportation for the journey. Looking again at Table \ref{table:desc}, we see that 44\% of the informed guests used a means of mass transportation, while, on the other hand, only 22\% of the non-informed guests traveled by public transportation. Moreover, among the treated guests, 41\% used the free arrival and departure offer to travel to the accommodation.

\begin{table}[H]
	\caption{Mean and standard deviation by information status}\label{table:desc}
	\begin{center}
		\begin{tabular}{lcc}
			\hline
			& \textbf{Informed guests ($D=1$)}                            & \textbf{Uninformed guests ($D=0$)}                        \\ \hline
			Hotel-specific ratio of informed guests & 0.61 (0.29) & 0.41 (0.27) \\
			Holiday flat             & 0.19 (0.39) & 0.25 (0.43) \\
			Train accessibility           & 0.91 (0.29) & 0.83 (0.38) \\
			Alone                    & 0.12 (0.33) & 0.11 (0.32) \\
			Family                    & 0.19 (0.40) & 0.25 (0.43) \\
			Purpose nature           & 0.63 (0.48) & 0.71 (0.46) \\
			Length of stay          & 4.75 (2.11) & 4.39 (1.85) \\
			Distance car           & 164.80 (74.98) & 169.73 (79.65) \\
			Travel time difference   & 89.59 (23.47) & 92.62 (23.59) \\
			Swiss residence   & 0.92 (0.27) & 0.89 (0.32) \\
			Car ownership           & 0.84 (0.37) & 0.85 (0.36) \\
			Half Fare Travelcard      & 0.82 (0.39) & 0.71 (0.46) \\
			Age                       & 60.73 (13.93) & 55.87 (14.66) \\
			Women                      & 0.56 (0.50) & 0.52 (0.50) \\
			High income              & 0.10 (0.29) & 0.09 (0.31) \\
			Public transport & 0.44 (0.50) & 0.22 (0.41) \\
			Free arrival-departure offer & 0.41 (0.49) & 0.00 (0.00) \\
			Observations            &              530                                      &                         124                            \\ \hline
				\end{tabular}
					\end{center}
		\begin{tablenotes}[flushleft]
			\footnotesize
			\item \textit{Notes: The sample contains guests who stay more than two nights in Appenzell Innerrhoden. }
		\end{tablenotes}
	\end{table}

\section{Results}\label{Results}

\subsection{Common support, match quality and random sample}\label{Common_quality_random}

Using matching methods, we assume that there are appropriate control observations to match with. According to Assumption 2, common support, there must be substantial overlap in the distribution of matching variables when comparing the treatment and control observations. Using statistical parlance, we must not be able to deterministically observe the treatment (i.e., information) status of an individual based on its covariates. Using the propensity score, we are obligated to observe a substantial overlap of the propensity score's $p(X)$ distribution, and none of the propensity scores should be zero or one. 

\begin{figure}[H] 
	\includegraphics[scale=1]{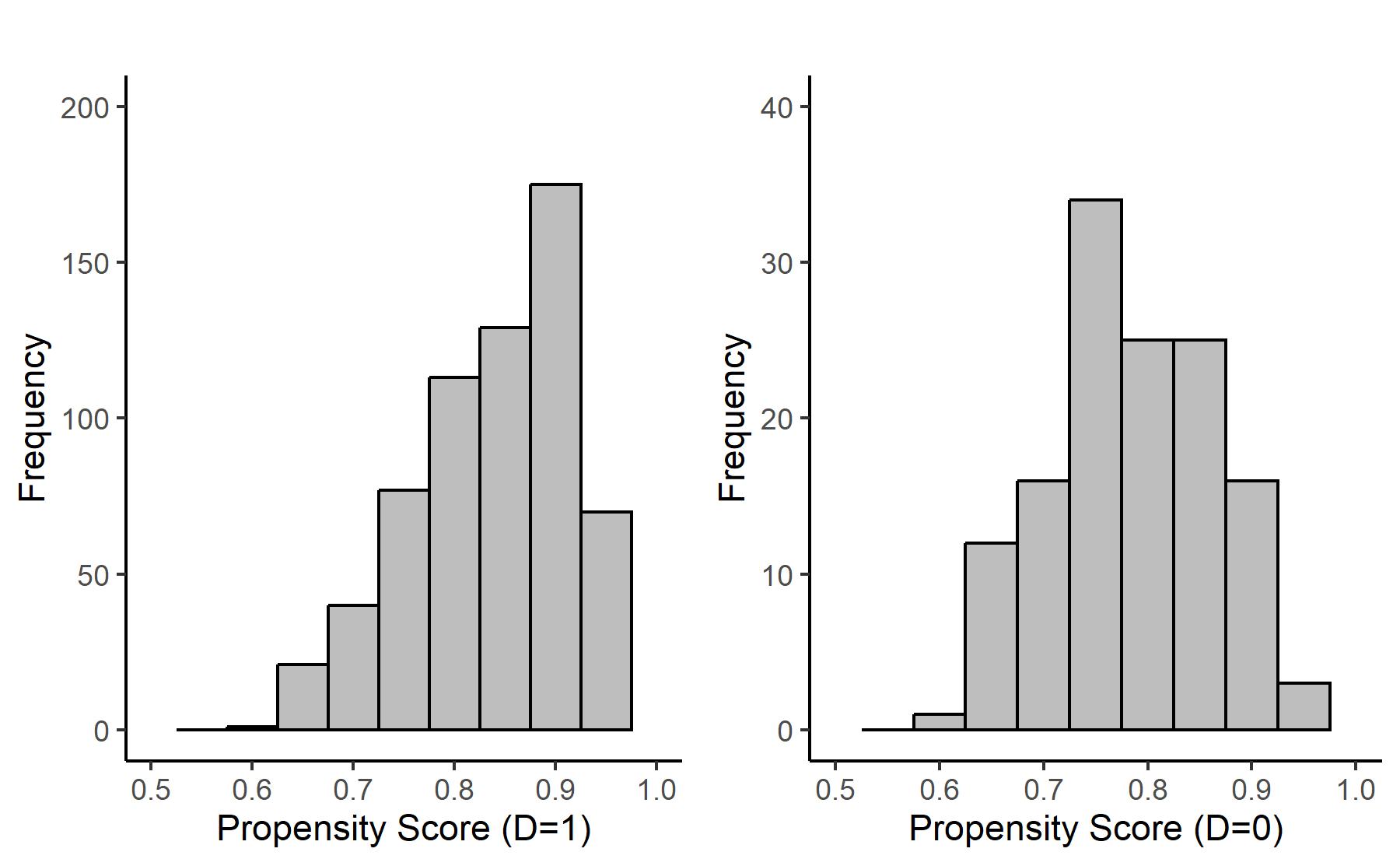}
	\centering \caption{Propensity scores of treatment and control group} \label{Fig:propscore}
\end{figure}

Looking at Figure \ref{Fig:propscore}, depicting the propensity scores of the causal forest estimation, we observe that Assumption 2 is fulfilled, as we observe a decent overlap: All observations have (at least a few) observations that have comparable propensity scores. Therefore, we can also assume that we have no major differences in unobserved characteristics. However, we also face some observations with relatively high propensity scores, i.e., scores close to one. Therefore, as a robustness check, we re-estimate the effect when trimming the propensity score (see Section \ref{Result:Robust}). 

The idea of matching methods is to compare the outcomes of informed and non-informed (about the free arrival and departure offer) individuals that are similar in terms of covariates, i.e., treatment and control groups are balanced. Table \ref{table:balance} presents the mean values of pre-selected variables that differ between treatment and control groups before propensity score matching, i.e., raw data. (The thing to notice is that we do not observe significant differences in the raw data regarding the travel composition (family), purpose (nature), or age.) We see that there exist differences at the 5\% significance level before matching for the variables "hotel-specific ratio of informed guests", train accessibility, and Half Fare Travelcard by looking at the p-values for a t-test, testing if the means are different in the treated and control groups. However, after matching, there are no meaningfully large (significant) differences in the means for the variables presented in Table \ref{table:balance}. Therefore, we conclude that treatment and control groups are balanced.  

\begin{table}[H]
		\caption{Balance table before and after matching}\label{table:balance}
		\begin{center}
	\begin{tabular}{lcc}
					\hline
		& \textbf{Before Matching} & \textbf{After Matching} \\			\hline
		\multicolumn{3}{l}{\textbf{Hotel-specific ratio of informed guests}} \\
		Mean Treatment  & 0.612                 & 0.576               \\
		Mean Control    & 0.407                 & 0.575              \\
		Std. Mean Diff  & 70.793                  & 0.558               \\
		t-test p-value  & \textless{}0.001        & 0.825               \\
		\multicolumn{3}{l}{\textbf{Train accessibility}}                            \\
		Mean Treatment  & 0.908                 & 0.899                  \\
		Mean Control    & 0.831                & 0.901               \\
		Std. Mean Diff  &  26.524                 & -0.761              \\
		t-test p-value  & 0.035                & 0.878                \\
		\multicolumn{3}{l}{\textbf{Half Fare Travelcard}}                            \\
		Mean Treatment  & 0.819                  & 0.790                 \\
		Mean Control    & 0.710                 & 0.821                \\
		Std. Mean Diff  & 28.325                 & -7.705                 \\
		t-test p-value  & 0.015                 & 0.110                 \\
		\hline
	\end{tabular}
\end{center}
\end{table}

\begin{figure}[H] 
	\includegraphics[scale=.2]{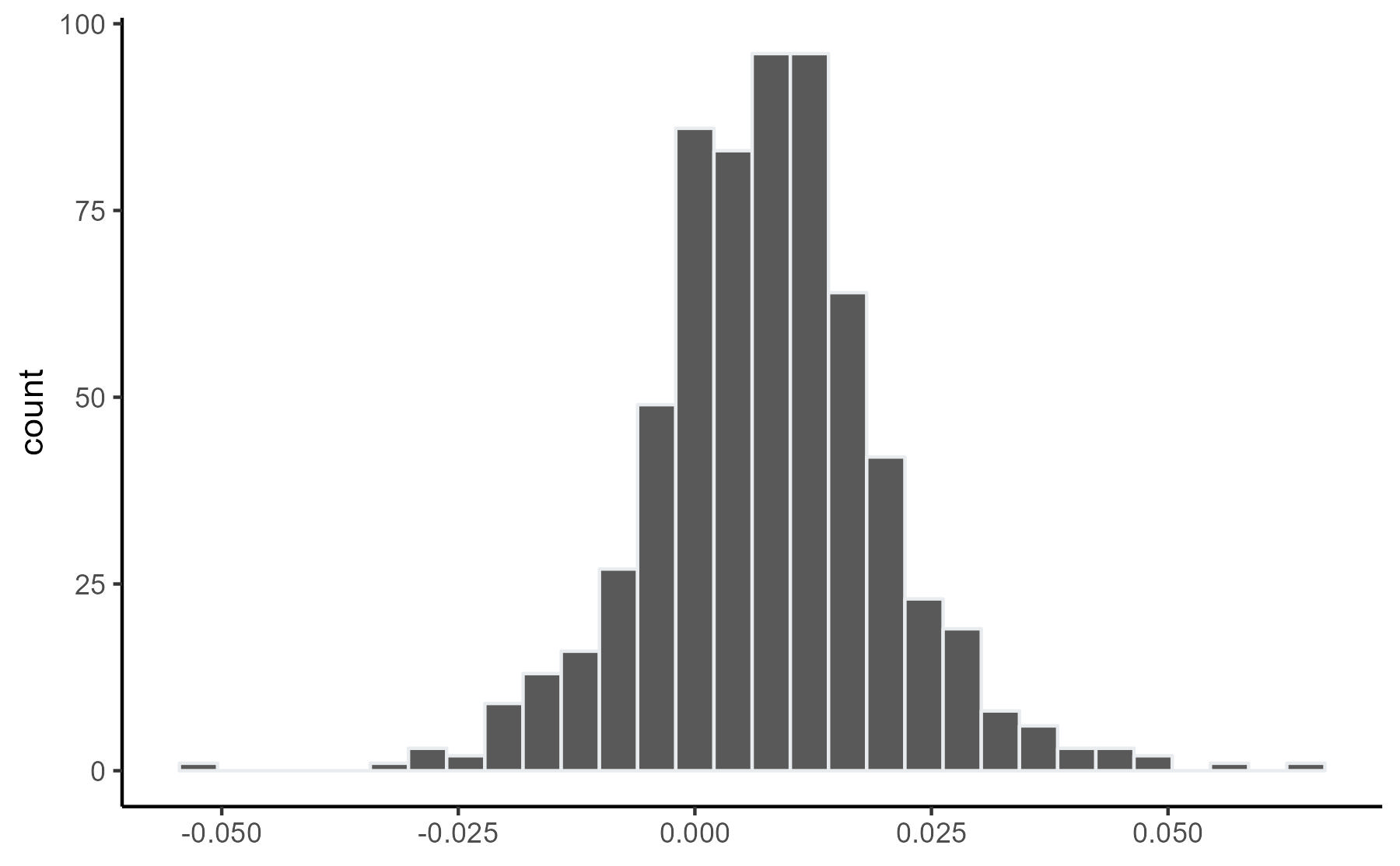}
	\centering \caption{Differences between the CATEs} \label{Fig:catediff}
\end{figure}

Finally, we check assumption 4 that our sample conditioning on $X$ is as good as if it were a random sample from the population of interest. To test whether guest card owners---the vast majority in our sample---differ from the overall population, we model the tree structure of the causal forest based on the original sample as well as on the subsample of all guest card owners. For both models, we then estimate the CATEs for all observations. If our sample conditioning on $X$ is as good as if it were a random sample, then the CATEs of the two tree structures should be similar. In Figure \ref{Fig:catediff}, we display the difference between the two estimated CATEs for each observation. The differences between the CATEs are minimal; however, they are significant and positive. Concluding, if we have a selection problem due to the sampling process, it is tiny and influences our results to a negligible extent. The thing to notice is that the causal framework (see Figure \ref{Fig:Defcausal}) also controls for the relevant covariates that jointly influence the ownership of the guest card.

\subsection{The effect of the free public transport offer}

Table \ref{table:eff1} shows the estimates of the free arrival and departure offer on travel mode choice, namely the main result of our analysis. When applying the causal forest, we obtain an average treatment effect (ATE) of 0.116, indicating that the information about the free arrival and departure offer, on average, increases the number of guests using public transportation by 11.6 percentage points. Considering the estimate of the propensity score matching, we arrive at an average treatment effect of 0.148, suggesting that the information increases the number of mode shifts towards public transportation by 14.8 percentage points. Both estimates are significant at the 5\% level (with the estimate of the causal forest being significant at the 1\% level). Hence, our estimates point to a positive average treatment effect of the free arrival and departure offer on travel mode choice. 

	\begin{table}[H]
	\caption{Effects on mode shift}\label{table:eff1}
	{\footnotesize
		\begin{center}
			\begin{tabular}{l c c }
				\hline
				& \textbf{Causal forest} &  \textbf{Propensity score matching} \\
				\hline
				Effect & 0.116 &  0.148 \\
				Standard error & 0.043 & 0.065 \\
				p-value & \textless{}0.001 &  0.023 \\
				\hline
				Number of observations & \multicolumn{2}{c}{654}\\
				\hline
			\end{tabular}
		\end{center}
		\begin{tablenotes}[flushleft]
			\footnotesize
			\item 
		\end{tablenotes}
		\par
	}
\end{table}

\begin{figure}[H] 
	\includegraphics[scale=.2]{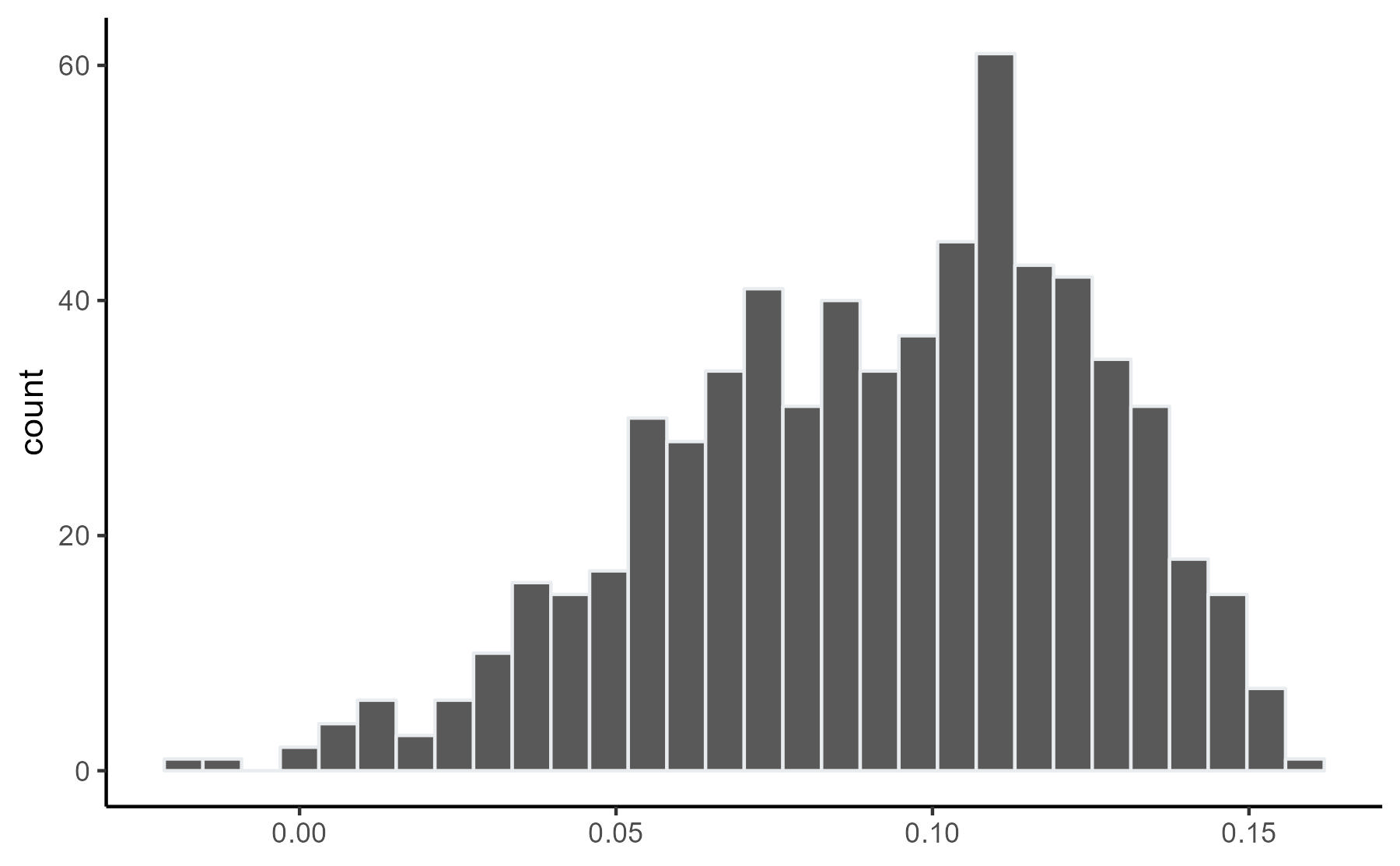}
	\centering \caption{Histogram of CATEs} \label{Fig:histCATEs}
\end{figure}

Considering the heterogeneity of the effects estimated by the causal forest in greater detail, Figure \ref{Fig:histCATEs} depicts the distribution of the conditional average treatment effects (CATEs). In conclusion, the CATEs are almost exclusively positive, which also shows the validity of the treatment effect among different guest groups.

\subsection{Robustness Checks}  \label{Result:Robust}

To challenge our estimates presented in the previous chapter, we obtain estimates of robustness investigations. 

As a first check, we challenge our results in a sense that we assume to have comparable propensity scores. This check stems from the fact that some observations with relatively high propensity scores are close to one. To do so, we i) change the target sample of the causal forest using the weighting scheme of \citet{li2018balancing}, in which each unit's weight is proportional to the probability of that unit being assigned to the opposite group. The thing to notice is that we now observe an average treatment effect for the overlap population (ATO).\footnote{To be precise, we set in the statistical software \textsf{R} the \texttt{target.sample}-argument from \texttt{target.sample="all"} to \texttt{target.sample="overlap"}.} Moreover, using propensity score matching, we ii) apply the trimming rule of \citet{dehejia1999causal} and omit all treatment group observations with a propensity score higher than the highest value among the control group.\footnote{Therefore, we change the \texttt{CommonSupport}-command to \texttt{TRUE} in the statistical software \textsf{R}.} 

Looking at Table \ref{table:robustness}, we see that, again, both estimates are positive and significant. The value of the propensity score matching is almost the same and only 0.7 percentage points higher. Also, the estimate of the causal forest is comparable to our original result (i.e., 2.5 percentage points lower).  

	\begin{table}[H]
	\caption{Effects on mode shift}\label{table:robustness}
	{\footnotesize
		\begin{center}
			\begin{tabular}{l c c c }
				\hline
				Check & 1 &  2 & 3 \\
				\hline
				\textbf{Causal Forest} & & & \\
				Effect & 0.091 &  0.119  & 0.205\\
				Standard error & 0.039 & 0.041 & 0.044 \\
				p-value & 0.020 & 0.003 &  0.000 \\
				\textbf{Propensity score matching} & & & \\
				Effect & 0.155 & 0.134 & 0.301 \\
				Standard error & 0.057 & 0.067 & 0.101 \\
				p-value & 0.009 & 0.018 & 0.003 \\
				\hline
				\textbf{Number of observations} & 654  & 843  & 583  \\ 
			\hline
			\end{tabular}
		\end{center}
		\begin{tablenotes}[flushleft]
			\footnotesize
			\item 
		\end{tablenotes}
		\par
	}
\end{table}

Second, we impute the missing data of covariates using multiple imputation to account for the uncertainty implemented by the MICE algorithm as described by \cite{van2011mice} and \cite{van2018flexible}.\footnote{We apply the \texttt{mice}-command of the package \texttt{mice} in the statistical software \textsf{R}.} So, we can include the 189 observations that have missing variables, from which 157 observations have only one covariate missing. The estimates are very similar to the main results, amounting to 0.134 for the propensity score matching and 0.119 for the causal forest matching. Therefore, we conclude that covariate missings are missings at random. 

Third, we replace our control group of uninformed guests with tourists who stay more than three nights in the cantons of Appenzell Ausserrhoden and Toggenburg. In Appendix \ref{Appendix_A}, we present the descriptive statistics for this robustness check. The thing to notice is that the control group becomes (too) small as we could not contact these guests by email (and thus, it might be that this control group is only in a limited sense comparable to the treated group). However, the impact of the treatment on demand shift is remarkable and amounts, depending on the algorithm, 0.205 and 0.301.

As we set out to learn something about free arrival and departure offers in general, our robustness checks further indicate that these have a meaningful causal effect on the choice of means of transport.

\section{Discussion}\label{Discussion}

We estimate a treatment effect of 11.6 and 14.8 percentage points. These estimates of the fare-free arrival and departure policy for overnight guests are comparable to the effects of the fare-free policy in Tallinn (14\%) and the fare-free subway policy for citizens aged 65 or above in Seoul (16\%), see \cite{cats2017prospects} and \cite{shin2021exploring}. Assuming that only overnight guests who ordered a free departure-arrival ticket changed their behavior due to the information provided by the hotelier (41.3 percent in the treatment group), we can calculate that 28.1\% (11.6/41.3) respective 35.8\% (14.8/41.3) of the overnight guests using the free arrival and departure offer would not arrived public transport in the absence of the free arrival and departure offer.

Ecologically of great relevance, we can again estimate the mode shifts influence on CO$_2$ emissions based on assumptions. Put simply, we assign a CO$_2$ value to the average routing distances per means of transport. Following, e.g., \citet{ohnmacht2020relationships}, we base our values on the so-called "mobitool factors"\footnote{See \href{https://www.mobitool.ch/de/tools/mobitool-faktoren-v3-0-25.html}{https://www.mobitool.ch/de/tools/mobitool-faktoren-v3-0-25.html} (accessed on October 24, 2023).} to assess the environmental impacts of different means of transport per person-kilometer. The CO$_2$ values include direct operation, vehicle maintenance, indirect CO$_2$ emissions caused by energy provision, vehicle manufacture, and the CO$_2$ emissions used for the infrastructure (track/road). The CO$_2$ values per person-kilometre for car (fleet average) and public transport (average public transport) amount to 186.4 and 12.4 gram CO$_2$ per person-kilometre, respectively. Therefore, using these two CO$_2$ values per distance and an average travel distance per means of transportation using Google Maps data (i.e., 165.8 for car and 187.7 for public transport, we can calculate an equivalent that reflects the CO$_2$ savings of the guests shifting transport mean. The savings amount to 57.2 kilograms CO$_2$ for every person traveling with public transport instead of a private car (for the calculation, see Appendix \ref{Appendix_CO2}). The Swiss mean of domestic CO$_2$ emission (equivalence) for transportation amounts to about 1.62 tons per person and year \citep[see for Swiss CO$_2$ emissions and population][]{UVEK_Kenngroessen_2023,BFS_Wohnbevoelkerung_2023}. Concluding, the usage of the offer (to and from the destination) reduces the yearly domestic CO$_2$ transportation emissions in Switzerland by 3.6\%. The share of domestic leisure travels would, therefore, be higher and the share of total transport, including international (air) travel, lower. 

Our results are, according to our robustness checks, valid for similar settings in which the targeted guest segments have a proneness towards public transport, and the quality of public transport services is high. Conversely, the external validity is limited for target guest groups with higher constraints (e.g., skiing tourism with more luggage to transport) and with a lack of quality in public transport services \citep[e.g., international travels across poorly connected subnetworks ][]{grolle2024service}. Hence, future studies (in other settings or with more statistical power) should investigate whether there is significant effect heterogeneity with respect to trip-related constraints \citep[see, e.g., ][]{huber2022business}. These insights on effect heterogeneity may then also support the elaboration of an optimal financial scheme policy for integrated products of public transport and accommodation.

\section{Conclusion}\label{Conclusion}

We assessed the causal effect of a free arrival and departure offer for overnight guests of a Swiss tourism destination. Based on the so-called "selection on observable" assumption, we take advantage of the random element that the information on the offer from a hotelier to a guest varies in day-to-day business. Using the causal forest and propensity score matching, we found that public transportation usage increases by 11.6 and 14.8 percentage points, depending on the method. The results also stand up to robustness checks, indicating that the effect lies between 9.1 and 15.5 percentage points. 

Our paper is the first to provide empirical evidence for researchers and policy-makers on how such a free arrival and departure offer influences the (domestic) guests' transport mode choice in Switzerland. Our estimands are essential for designing future comparable offers in light of CO$_2$ reductions, as a shift towards public transport helps decrease CO$_2$ emissions. To this end, our empirical approach may also be applied to comparable natural experiments of the travel and tourism industry in Switzerland or other countries. 
	
	\newpage
	\bigskip
	
	\bibliographystyle{econometrica}
	\bibliography{Appenzell.bib}
	
	\newpage
	\bigskip

\begin{appendix}
		
		\numberwithin{equation}{section}
		\counterwithin{figure}{section}
		\noindent \textbf{\LARGE Appendices}
	
\section{Causal framework including the propensity score}\label{Appendix_Causal_Framework_PSM}

\begin{figure}[H]
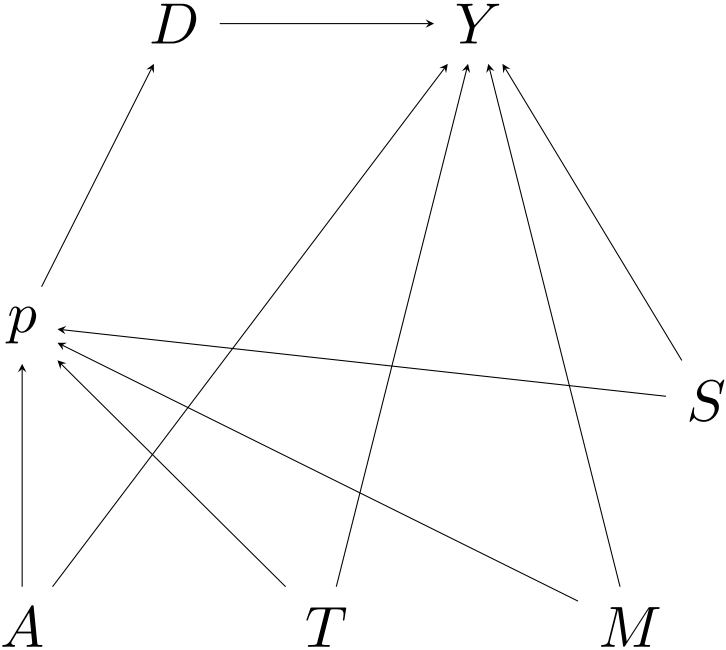

	\centering \caption{\label{figuresetup_PSM}  Causal framework including the propensity score (denoted by p)}\bigskip
	\definecolor{offwhite}{HTML}{F2EDED} 
	\tikzset{> = stealth} 
	\tikz{ 
		\node[black, scale=2] (d) at (2,8) {$D$}; 
		\node[black, scale=2] (p) at (0,4) {$p$}; 
		\node[black, scale=2] (t) at (0,0) {$A$}; 
		\node[black, scale=2] (s) at (4,0) {$T$}; 
		\node[black, scale=2] (m) at (8,0) {$M$};
		\node[black, scale=2] (a) at (9,3) {$S$};
		\node[black, scale=2] (y) at (6,8) {$Y$}; 
		\path[->, draw=black] (p) edge (d); 
		\path[->, draw=black] (d) edge (y); 
		\path[->, draw=black] (t) edge (y); 
		\path[->, draw=black] (s) edge (y); 
		\path[->, draw=black] (m) edge (y);
		\path[->, draw=black] (t) edge (p); 
		\path[->, draw=black] (s) edge (p); 
		\path[->, draw=black] (m) edge (p);
		\path[->, draw=black] (a) edge (p);
		\path[->, draw=black] (a) edge (y);
		
	}
\end{figure}

	\section{Descriptive statistics robustness checks}\label{Appendix_A}

\begin{table}[H]
	\caption{Descriptive statistics robustness check 2: Data imputation}\label{table:desc_imputation}
	\begin{center}
		\begin{tabular}{lcc}
			\hline
			& \textbf{Informed guests ($D=1$)}                            & \textbf{Uninformed guests ($D=0$)}                        \\ \hline
			Hotel-specific ratio of informed guests & 0.61 (0.29) & 0.41 (0.27) \\
			Holiday flat             & 0.18 (0.39) & 0.28 (0.45) \\
			Train accessibility           &  0.91 (0.28) & 0.85 (0.36) \\
			Alone                    & 0.13 (0.34) & 0.10 (0.30) \\
			Family                    & 0.20 (0.40) & 0.25 (0.43) \\
			Purpose nature           & 0.62 (0.49) & 0.72 (0.45) \\
			Length of stay          & 4.71 (2.08) & 4.38 (1.83) \\
			Distance car           & 163.47 (75.93) & 163.64 (79.95) \\
			Travel time difference   & 89.74 (23.53) & 92.26 (23.43) \\
			Swiss residence   & 0.91 (0.29) & 0.87 (0.34) \\
			Car ownership           & 0.86 (0.35) & 0.83 (0.37) \\
			Half Fare Travelcard      & 0.80  (0.40) & 0.69 (0.46) \\
			Age                       & 61.30 (13.84) & 56.15 (14.47) \\
			Women                      & 0.56 (0.50) & 0.53 (0.50) \\
			High income              & 0.09 (0.28) &  0.09 (0.29) \\
			Public transport & 0.41 (0.49) & 0.22 (0.42) \\
			Free arrival-departure offer &  0.39 (0.49) & 0.00 (0.00) \\
			Observations            &                  687                                &                             156                       \\ \hline
				\end{tabular}
					\end{center}
		\begin{tablenotes}[flushleft]
			\footnotesize
			\item \textit{Notes: The sample contains guests who stay more than two nights in Appenzell Innerrhoden. }
		\end{tablenotes}
	\end{table}

	\begin{table}[H]
		\caption{Descriptive statistics robustness check 3: Appenzell Ausserrhoden and Toggenburg}\label{table:desc:Ausserroden}
		\begin{center}
			\begin{tabular}{lcc}
			\hline
			& \textbf{Informed guests ($D=1$)}                            & \textbf{Control group}                        \\ \hline
			
			Holiday flat             & 0.19 (0.39) & 0.38 (0.49) \\
			Train accessibility           & 0.91 (0.29) & 0.66 (0.48) \\
			Alone                    & 0.12 (0.33) & 0.09 (0.30) \\
			Family                    & 0.19 (0.40) & 0.38 (0.49) \\
			Purpose nature           & 0.63 (0.48) & 0.81 (0.39) \\
			Length of stay          & 4.75 (2.11) & 5.30 (2.36) \\
			Distance car           & 164.80 (74.98) & 169.73 (79.65) \\
			Travel time difference   & 89.59 (23.47) & 92.62 (23.59) \\
			Swiss residence   & 0.92 (0.27) & 0.89 (0.32) \\
			Car ownership           & 0.84 (0.37) & 0.79 (0.41) \\
			Half Fare Travelcard      & 0.81 (0.39) & 0.74 (0.45) \\
			Age                       & 60.73 (13.93) & 52.62 (13.26) \\
			Women                      & 0.56 (0.50) & 0.45 (0.50) \\
			High income              & 0.09 (0.29) & 0.11 (0.32) \\
			Public transport & 0.44 (0.50) & 0.25 (0.43) \\
			Free arrival-departure offer & 0.41 (0.49) & 0.00 (0.00) \\
			Observations & 530&53   \\ \hline
			\end{tabular}
		\end{center}
		\begin{tablenotes}[flushleft]
			\footnotesize
			\item \textit{Notes: The sample contains guests who stay more than two nights in Appenzell Innerrhoden (treatment group), Appenzell Ausserrhoden, and Toggenburg (both control group). }
		\end{tablenotes}
	\end{table}

		\section{Calculation of the CO$_2$ emissions}\label{Appendix_CO2}
	
	Savings for a person using public transport instead of private car to and from a destination: 
	
	\begin{equation} 2*\frac{165.8*186.4-187.7*12.4}{1000} = 57.2 \text{ kilogram CO$_2$} \end{equation}

	\end{appendix}
\end{document}